\documentclass[aps,prd,preprintnumbers,showpacs]{revtex4}
\usepackage[dvips]{graphicx}
\begin{document}

%
%

\eprint{Nisho-2-2017}
\title{Chiral Symmetry Breaking by Monopole Condensation}
\author{Aiichi Iwazaki}
\affiliation{International Economics and Politics, Nishogakusha University,\\ 
6-16 3-bantyo Chiyoda Tokyo 102-8336, Japan.}   
\date{Feb. 1, 2017}
\begin{abstract}
Under the assumption of Abelian dominance in QCD,
we have shown that chiral condensate is locally present around each QCD monopole. 
The essence is that either charge or chirality of a quark is not conserved, when the low energy massless quark collides with QCD monopole.
In reality, the charge is conserved so that the chirality is not conserved.  
Reviewing the presence of the local chiral condensate,
we show by using chiral anomaly that chiral non symmetric quark pair production takes place when a color charge is putted in a vacuum
with monopole condensation, 
while chiral symmetric pair production takes place in a vacuum with no monopole condensation.  
Our results strongly indicate that the chiral symmetry is broken by the monopole condensation.
\end{abstract}
\hspace*{0.3cm}

\hspace*{1cm}

\maketitle

\section{introduction}
It has been shown with lattice gauge theories\cite{karsch} that quark confinement and chiral symmetry breaking simultaneously arises in SU(3) gauge theory
with massless quark color triplets.
That is, the transition temperature between confinement and deconfinement phases 
almost coincides with the transition temperature 
between chiral symmetric and antisymmetric phases. Although extensive studies\cite{suga,miyamura,w,fuku,gat,sch,iritani} have been performed,
the explicit connection between the confinement and the chiral symmetry breaking has been still not
clear.  
The confinement
is caused by the monopole condensation\cite{nambu,man,thooft} in the analysis with the use of maximal Abelian gauge\cite{mag}. 
On the other hand, the chiral symmetry
breaking is caused by the chiral condensation of quark-antiquark pair. It has been discussed to arise with 
instanton effects through chiral anomaly. No theoretical relation between
the monopole condensate and the chiral condensate was found, although there were 
numerical evidences\cite{miyamura} that the monopole condensates are
correlated with the chiral condensates.

We have recently discussed\cite{iwazaki} the breakdown of the chiral symmetry by the monopole condensation.
The point is the presence of the chiral condensate around each monopole. Such a presence was originally shown
in the analysis of Rubakov effects\cite{rubakov,callan,ezawa}; baryon number non conservation caused by the collisions between nucleons and
GUT monopoles ( 'tHooft-Polyakov monopoles\cite{mono} ).
Although the purpose of the analysis was to show the baryon number non conservation in the baryon scattering with the monopoles,
the analysis reveals the presence of the chiral condensate as well as baryon number condensate
around the monopoles.
We have applied the fermion monopole dynamics to the quark scattering with QCD monopoles 
using the assumption of the Abelian dominance\cite{iwa,suzuki}.
The Abelian dominance dictates that the low energy phenomena such as quark confinement and chiral symmetry breaking are
described by the massless gauge fields ( diagonal gluons ) of maximal Abelian gauge groups, QCD monopoles and massless quarks.
The fermion monopole dynamics in the Rubakov effects is also described by the massless Abelian gauge fields, e.g. electromagnetic fields,  
GUT monopoles, quarks and leptons. Heavy gauge fields and Higgs fields in GUT are irrelevant to the phenomena
since we are only concerned with the phenomena of much lower energies than those of GUT.
The situation is similar to the case of the Abelian dominance. That is, off diagonal gluons are massive and irrelevant to the
low energy phenomena in QCD. 
Therefore, we have shown the local presence around each QCD monopole by using
the similarity between the quarks coupled with the QCD monopoles and the fermions coupled with GUT monopoles.
The essence in the monopole fermion dynamics is that in the fermion scattering with a monopole, the fermions much satisfy either 
chiral non conserved but charge conserved, or charge non conserved but chirality conserved boundary condition 
at the location of the monopole; the charge is associated with
the relevant Abelian gauge group.
In reality, the charge conserved but chirality non conserved boundary condition should be satisfied. 
The fact leads to the local chiral condensate around the monopole.
The presence of the local chiral condensate around QCD monopole is 
indicative of the chiral symmetry breaking by the monopole condensation.

\vspace{0.1cm}
In this paper we show using chiral anomaly that when an external charge is put in a vacuum with the monopole condensation,
the vacuum expectation value $\langle dQ_5/dt\rangle $ 
does not vanish,
while it vanishes when the vacuum has no monopole condensation. Here $dQ_5/dt$ denotes the time derivative of the chirality $Q_5$.
That is, it represents the pair production of the chiral fermions under the background electric field of the external charge. 
We note that $Q_5=N_R-N_L$ where $N_R$ ( $N_L$ ) denotes the number of the right ( left ) handed fermion.  
Therefore, the chiral non symmetric spontaneous pair production of the massless fermions takes place in the vacuum with monopole condensation.
The result strongly indicates that the chiral symmetry is broken by the monopole condensation. 
( In general, the pair production of the massless fermions under an electric field is chiral symmetric when there is no magnetic field. The positive charged fermion
produced moves to the direction of the electric field, while the negative charged fermion does to the opposite direction of the electric field.
Their spins can take arbitrary directions parallel or anti parallel to the electric field. There are no favorable directions. Therefore, the production
is chiral symmetric; $Q_5=0$. )

In the next section(\ref{22}), we explain why either the charge or the chirality of a fermion is not conserved 
in the low energy massless fermion scattering with a Dirac monopole.
We find that an appropriate boundary condition for the fermion must be imposed in order to define the fermion scattering with the monopole. 
In the section(\ref{33}), by solving Dirac equation, we discuss the appropriate boundary condition imposed at the location of the monopole.
We find that it is uniquely determined in the case of 'tHooft-Polyakov monopole, while it is not in the case of QCD monopole.
Dirac monopoles represent these monopoles apart from their cores.
In the section(\ref{44}), we explain physically why the real boundary condition is charge conserved but chirality non conserved one.
We also show that there is chiral condensate locally present around both 'tHooft-Polyakov and QCD monopoles.
In the section(\ref{55}) we show our main result that chiral non symmetric pair production arises when an external charge is put in a vacuum with monopole condensation.
We summarize and discuss our results in the final section(\ref{66}). 

\section{either charge or chirality conservation}
\label{22}  	
First, we explain that either the charge or the chirality of a fermion is not conserved in the low energy massless fermion scattering with a Dirac monopole.
We need to impose an appropriate boundary condition on the fermion at the location of the monopole, which
determines the conserved quantity.

Our concern is 
massless fermion doublet $(\begin{array}{l}\Psi_+ \\ \Psi_-\end{array} )$ satisfying Dirac equation in the background gauge fields of the monopole,

\begin{equation}
\label{1}
\gamma_{\mu}(i\partial^{\mu}\mp \frac{g}{2}A^{\mu})\Psi_{\pm}=0
\end{equation}
where the gauge potentials $A^{\mu}$ denotes a Dirac monopole
given by 

\begin{equation}
\label{2}
A_{\phi}=g_m(1-\cos(\theta)), \quad A_0=A_r=A_{\theta}=0
\end{equation} 
where $\vec{A}\cdot d\vec{x}=A_rdr+A_{\theta}d\theta+A_{\phi}d\phi$ with polar coordinates $r,\theta$ and $\phi=\arctan(y/x)$.
$g_m$ denotes a magnetic charge with which magnetic field is given by $\vec{B}=g_m\vec{r}/r^3$.
The magnetic charge satisfies the Dirac quantization condition $g_mg=n/2$ with integer $n$ where $g$ denotes the U(1) gauge coupling.
Hereafter, we assume the monopoles with the magnetic charge $g_m=1/2g$.

\vspace{0.1cm}
The fermion doublet coupled with the monopole arises in gauge theories, 
for example, quarks in SU(2) gauge theories under the assumption of the Abelian dominance\cite{iwa,suzuki}.
The maximal Abelian group is described by the diagonal component ( $\sigma_3$ ) of the SU(2) gauge fields. 
The Dirac monopole is represented by the Abelian gauge fields. 
Thus, the quark doublet $q=(q^+,q^-)$ carry the charges $\pm g/2$ associated with the diagonal component.
Similarly, quark doublets coupled with the monopoles in SU(3) gauge theories arises under the assumption of the Abelian dominance.
The maximal Abelian group is described by the diagonal components of gluons; $A_{\mu}^3$ and $A_{\mu}^8$.
In SU(3) gauge theory, we have three types of monopoles\cite{coleman}, which are characterized by root vectors of SU(3),
 $\vec{\epsilon}_1=(1,0), \vec{\epsilon}_2=(-1/2,-\sqrt{3}/2)$ and $\vec{\epsilon}_3=(-1/2,\sqrt{3}/2)$.
They describe the couplings with the maximal Abelian gauge fields, $A_{\mu}^{3,8}$ such as $\epsilon_i^a A_{\mu}^a$.
For example a monopole characterized by $\vec{\epsilon}_1$ produces the magnetic fields represented by $A_{\mu}^3=\epsilon_1^a A_{\mu}^a$. 
Thus, the quarks coupled with the monopole
are a doublet $q=(q^+,q^-,0)$ of the color triplet. Similarly the other monopoles couple with the quark doublets, $q=(q^+,0,q^-)$ and $q=(0,q^+,q^-)$. 
Thus, there are three types of the quark doublets.

On the other hand, SU(2) doublet was originally considered in the discussion of the Rubakov effects\cite{rubakov,callan,ezawa} 
where SU(2) gauge group embedded in unification groups, e.g. SU(5), is broken into U(1) gauge group.
The monopole solitons (  'tHooft-Polyakov monopoles ) appear when
the gauge group SU(2) is broken into U(1) gauge group by a triplet Higgs.
The fermion doublet coupled with the monopole is composed of quarks and leptons.
The field configurations of the  'tHooft-Polyakov monopoles are identical to those of the Dirac monopoles apart from their small cores.
In this way, the scattering of the fermion doublets with the monopoles in the Abelian gauge theory is important issue to understand
the low energy phenomena in GUT or QCD.
As we show below, either the charge or the chirality conservation is lost in the scattering.

\vspace{0.1cm}  

Now, we explain why either the charge or the chirality is not conserved when the fermion collides with the monopole. 
As is well known, the conserved angular momentum\cite{coleman} of the fermion under the magnetic monopole located at $r=0$ 
is given by $\vec{J}=\vec{L}+\vec{S}\mp gg_m\vec{r}/r$,
where $\vec{L}$ ( $\vec{S}$ ) denotes orbital ( spin ) angular momentum. The last term is peculiar to the particle under the background field
of the monopole. Owing to the term we can show that either the charge or the chirality is not conserved in the scattering.
In order to see it we note the conserved quantity $\vec{J}\cdot\vec{r}=\vec{S}\cdot\vec{r}\pm gg_m r$.
When the chirality ( or helicity $\sim \vec{p}\cdot\vec{S}/|\vec{p}||\vec{S}|$ ) is conserved, the spin must flip $\vec{S}\to -\vec{S}$ after the scattering 
because the momentum flips after the scattering; $\vec{p}\to -\vec{p}$.
Then, the charge must flip $g\to -g$ because of the conservation of $\vec{J}\cdot\vec{r}$, i.e.
 $\Delta(\vec{J}\cdot\vec{r})=\Delta(\vec{S}\cdot\vec{r})+\Delta(gg_mr)=0$. ( $\Delta(Q)$ denotes the change of the value $Q$ after the scattering. ) 
On the other hand, when the charge is conserved ( $0=\Delta(\vec{J}\cdot\vec{r})=\Delta(\vec{S}\cdot\vec{r}$) ),
the chirality $\vec{p}\cdot\vec{S}/|\vec{p}||\vec{S}|$ must flip because the spin does not flip $\vec{S}\to \vec{S}$.
In this way, either the charge or the chirality conservation is lost in the scattering.

We explain the fact in a different point of view. The lowest energy states under magnetic field
are those with their magnetic moments parallel to the magnetic field. When the low energy incoming fermion has the charge $g>0$,
the spin $\vec{S}$ must be parallel to $\vec{r}$, since the magnetic field $\vec{B}$ of the monopole is given by $\vec{B}\propto \vec{r}/r^3$. 
If the charge is conserved after the scattering, the spin of the outgoing fermion
must be also parallel to $\vec{r}$. Obviously the chirality $\vec{p}\cdot\vec{S}/|\vec{p}||\vec{S}|$ is not conserved because the momentum $\vec{p}$ 
flips after the scattering. On the other hand, if the charge flips $g\to -g$ after the scattering,
the spin of the outgoing fermion is anti-parallel to $\vec{r}$. Then, the chirality is conserved.
Therefore, either the charge or the chirality conservation does not hold in the scattering.

The presence of the magnetic monopole is essential in the discussion. 
We give an additional explanation using chiral anomaly.
The anomaly is particularly effective in the presence of the magnetic monopole.
As we describe later, the discussion just below is applied to show the chiral non symmetric pair production
i.e. $\langle dQ_5/dt \rangle\neq 0$ in the vacuum with the monopole condensation.

The anomaly equation describing the evolution of the chirality $Q_5$ is given by

\begin{equation}
\label{3}
\frac{dQ_5}{dt}=cg^2\int d^3r \vec{E}\cdot\vec{B}=cg^2\int d^3r \vec{E}\cdot\frac{g_m\vec{r}}{r^3}
=cg^2\int d^3r \frac{g(\vec{r}-\vec{x}(t))}{4\pi|\vec{r}-\vec{x}(t)|^3}\cdot\frac{g_m\vec{r}}{r^3}=\frac{c g^3g_m}{|\vec{x}(t)|}
\end{equation}
with the numerical constant $c$, 
where the electric field $\vec{E}=g(\vec{r}-\vec{x}(t))/(4\pi|\vec{r}-\vec{x}(t)|^3)$ is produced by a charged fermion located at $\vec{x}(t)$.
The anomaly equation describes how the chirality of the charged fermion changes with time $t$, depending on its coordinate $\vec{x}(t)$. 
Here, we consider the scattering of the charged fermion such that
the incoming fermion for $t<0$ approaches at the monopole and arrives the nearest point $\vec{x}_0\neq 0$ to the monopole at $t=0$, then it goes out for $t>0$. 
Namely, it goes from $\vec{x}(t=-\infty)=-\infty $ to $\vec{x}(t=\infty)=+\infty$,
passing $\vec{x}(t=0)=\vec{x}_0\neq 0$ at $t=0$. We assume that the coordinate $|\vec{x}(t)|$ is symmetric as $t\to -t$.
Then, when the fermion does not flip its charge after the scattering, the quantity $dQ_5/dt$ does not change its sign. Thus,
$Q_5(+\infty)-Q_5(-\infty)=\int_{-\infty}^{+\infty}dt\, dQ_5/dt$ does not vanish. 
The chirality is not conserved, when the charge is conserved.
On the other hand, 
when the charge flips $g\to -g$ after the scattering, $Q_5(+\infty)-Q_5(-\infty)=0$
because $dQ_5/dt$ change its sign after the scattering; $\vec{E}\to -\vec{E}$ for $t>0$. 
Therefore, we find that either the charge or the chirality conservation does not hold in the scattering.

\vspace{0.1cm}
It is important to make a comment on the anomaly equation(\ref{3}). As we have previously shown\cite{iwazaki2}, 
the anomaly equation describes the spontaneous production of charged massless
fermions under the electric field $\vec{E}$ and the magnetic field $\vec{B}$. 
( The idea is generally accepted. Especially, the observation of the pair production of chiral fermionic excitations has recently been reported in a semimetal\cite{semimetal}. ) 
In the present calculation,
the electric field $\vec{E}$ is produced by a charged fermion put at $\vec{x}=\vec{x}(t)$ and the magnetic field $\vec{B}$ is given by the monopole. 
Because $Q_5=N_R-N_L\neq 0$, the pair production is chiral non symmetric.
The right handed fermions are mainly produced.
Thus, the chirality is not preserved when the charge $g$ of the fermion 
is preserved in the scattering. 
We should remember that the pair production of the massless charged fermions in general arise in background electric fields.
In the case, both of right and left handed fermions are equally produced. Thus, the chirality is conserved.
On the other hand, when magnetic field is present in addition to the electric field, the chirality is broken because
the magnetic moments of the fermions are aligned by the magnetic field. Right ( or left handed ) fermions are more produced than
the left ( right ) handed fermions.

\section{boundary conditions}
\label{33}

The above arguments do not determine which one should be conserved, charge or chirality in the fermion scattering with the monopole.
We expect to obtain an appropriate boundary condition at $r=0$ by
solving the Dirac equation.

Thus,
we solve the Dirac equation(\ref{1}) with  $gg_m=1/2$. 
We take only the states of $\vec{J}=\vec{L}=0$, because the states approach the monopole at $r=0$.
Then, the solutions can be found in the following \cite{ezawa},

We set
\begin{equation}
\label{4}
\Psi_{\pm}=\frac{1}{r}\left(\begin{array}{l}f_{\pm} (r,t) \\ \mp ig_{\pm}(r,t)\end{array}\right )\eta_{\pm} \quad \mbox{with} \quad
\frac{\sigma_i x_i}{r}\eta_{\pm}=\pm \eta_{\pm}.
\end{equation}

Then the equation is decomposed into two independent equations,

\begin{equation}
\label{5}
i\bar{\gamma}_{\alpha}\partial^{\alpha}\psi_{\pm}=0 \quad \mbox{with} \quad \psi_{\pm}\equiv\left(\begin{array}{l}f_{\pm} (r,t) \\  -ig_{\pm}(r,t)\end{array}\right) 
\end{equation}
with $\alpha=0,1$, $x_0=t$ and $x_1=r$, where two dimensional gamma matrices are defined by

\begin{equation}
\label{6}
\bar{\gamma}^0=\left(\begin{array}{rr} 1 & 0 \\ 0  &-1 \\ \end{array}\right), \quad \bar{\gamma}^1=\left(\begin{array}{rr} 0 & 1 \\ -1  & 0 \\ \end{array}\right).
\end{equation}
We can easily solve the two dimensional equations (\ref{5}). The solutions are characterized by their chiralities and charges as well as  
their motions i.e. incoming or outgoing.  
When $E>0$, the solutions $\psi_+$ ( $\psi_-$ ) with positive charge ( negative charge ) is given by
\begin{eqnarray}
\psi_{\pm,in} &=&\exp(-iE(t+r))\left(\begin{array}{l}1 \\ -1\end{array}\right )\,; \quad \mbox{incoming fermions with left handed ($+$) and right handed ($-$)} \nonumber \\ 
\psi_{\pm,out} &=&\exp(-iE(t-r))\left(\begin{array}{l}1 \\ 1\end{array}\right )\,; \quad \mbox{outgoing fermions with right handed ($+$) and left handed ($-$)},
\end{eqnarray}
where the term ``incoming fermion with left handed ($+$)'' implies that incoming fermion is left handed with its charge being positive ($+$). 
Similarly, the term ``outgoing fermion with left handed ($-$)'' implies that outgoing fermion is left handed with its charge being negative ($-$).  
We can see from these solutions that either the charge or the chirality conservation is broken in the scattering.
For example, when an incoming fermion $\psi_{+,in}$ has a positive charge, its chirality is inevitably left handed.
( the magnetic moment must be parallel to the magnetic field. ) 
After the scattering, the outgoing fermion $\psi_{+,out}$ is right handed if
the charge is conserved. On the other hand, the outgoing fermion $\psi_{-,out}$ must have a negative charge if the chirality is conserved.
The fermion scattering is not uniquely defined in the system.

\vspace{0.1cm}
The non-uniqueness of the scattering comes from the fact that the Dirac monopole is singular at $r=0$.
It is well known\cite{boundary} that we need to impose a boundary condition at $r=0$ to define the fermion scattering with the Dirac monopole.
On the other hand, we are automatically led to choose a boundary condition in the fermion scattering with 'tHooft-Polyakov monopole regular at $r=0$.
The boundary condition is the chirality conserved but charge non conserved one.
Actually we can easily see the boundary condition.
Dirac equation with the 'tHooft-Polyakov monopole is given by

\begin{equation}
\label{8}
\gamma_{\mu}(i\partial^{\mu}-gA^{\mu})\Psi=0
\end{equation}
where the gauge potentials $A^{\mu}=A^{\mu}_a\sigma_a/2 $ denotes the field configuration of the 'tHooft-Polyakov monopole
( $\sigma_a$ denotes Pauli matrices ), 

\begin{equation}
A^{0}_a=0, \quad A^{i}_a=\epsilon_{a i j}\frac{x^j F(r)}{g r^2}.
\end{equation}
with indices $a=1,2,3$ and radial coordinate $r$. 

When $F\equiv1$, $A^{\mu}_a$ represents Wu-Yang monopole, a solution of Yang-Mills equations, which is singular at the origin $r=0$.
The equation is reduced to the Dirac equation (\ref{1}) when we take a gauge in which $A_a^i\sigma_a$ is transformed to $A_3'^i\sigma_3$; $A_3'^i$ describe
the Dirac monopole in eq(\ref{2}).
On the other hand, when the gauge group SU(2) is 
broken into the U(1) gauge group with a Higgs triplet,
a regular monopole solution arises with the smooth function $F(r)$;
$F(r)=1$ for $r>r_c$ and $F(r)\simeq 0$ for $r<r_c$; 
$r_c$ denotes the small radius of the monopole core, e.g. $r_c^{-1}\simeq 10^{15}$GeV.
The solution is called as  'tHooft-Polyakov monopole
regular at $r=0$ because of $F(r=0)=0$. 
Apart from the core, the equation is reduced to the Dirac equation (\ref{1}). 
But taking into account of the monopole core in the system,
we are forced to choose a boundary condition which conserves the chirality but breaks the charge conservation.
Namely, the boundary condition in the case of  'tHooft-Polyakov monopole, is given by $\psi_+(r=0)=\bar{\gamma}_0\psi_-(r=0)$
with the notation eq(\ref{5}) and eq(\ref{6}). 
Actually, the condition is derived by examining the equation valid even in the core,
$i\bar{\gamma}_{\nu}\partial^{\nu}\eta+(\frac{1-F}{r})\eta=0$ with $\eta\equiv (\psi_{+}-\bar{\gamma}_0\psi_{-})$ where
the last term vanishes outside the core of the monopole but it remains inside the core.
In realistic models involving the 'tHooft-Polyakov monopoles, the boundary condition also breaks baryon number conservation. 
Thus, the Rubakov effect comes out in the baryon scattering with the 'tHooft-Polyakov monopoles owing to the boundary condition. 
In this way, the fermion scattering with the 'tHooft-Polyakov monopoles can be uniquely defined.

\vspace{0.1cm}
A comment is in order.
It seems apparently that the U(1) gauge symmetry is broken by the boundary condition. But  we should note that
in the core of the monopole the gauge symmetry of SU(2) is restored because triplet Higgs field composing the monopole solutions 
vanishes in the core; the Higgs field breaks the gauge symmetry.
Thus, we can take a gauge in the core which transforms the positive charged fermion into the negative charged one.
Therefore, the charge non conserved boundary condition can be imposed without the breaking of the SU(2) gauge symmetry. However,
in compensation for the boundary condition, a charge is deposited on the monopole after the scattering.  
Namely, after the fermion scattering with the monopole, the monopole is charged; the monopole becomes a dyon.
But the production of the dyon is energetically unfavorable when we consider the low energy scattering.
In reality, the charge conservation should hold even if we take the charge non conserved boundary condition, as we explain the detail below.  

\vspace{0.1cm}
When we consider the quark doublets under the assumption of the Abelian dominance, 
there are no way to choose an appropriate boundary condition at $r=0$
because of the singularity at $r=0$. Because it is concerned with physics at short distance, 
the assumption of the Abelian dominance does not hold.
We must analyze original gauge theories themselves, e.g. color SU(3) gauge theory. At the short distance, the original gauge symmetry is effective.
Then, the quarks may change their charges in the vicinity of the monopole by the gauge transformation.
Thus, it is possible to have the charge non conserved but chirality conserved boundary condition.
In such a case, off diagonal gluons are produced in the scattering just as dyons are produced in the scattering with the 'tHooft-Polyakov monopole.
On the other hand, it is possible to have the chirality non conserved but charge conserved boundary condition.
In any case, we have two possible boundary conditions in QCD. The dynamics determines an appropriate boundary condition
as we briefly explain below.
     
\section{chiral condensate around a monopole}
\label{44}
When we take the boundary condition $\psi_+(r=0)=\bar{\gamma}_0\psi_-(r=0)$, which conserves the chirality but does not conserve
the charge, we can show that the charge conservation is restored but chirality conservation is broken. The restoration of the charge conservation
and the breakdown of the chirality conservation come from
the quantum effects of the Abelian gauge fields. This was shown previously in the analysis of the Rubakov effects.
Actually, taking the quantum fluctuations of the gauge fields $\delta A_{\mu}$, the Lagrangian describing the fermion scattering
with the monopole is given by

\begin{equation}
\label{10}
L=\bar{\psi}_+\bar{\gamma}_{\alpha}(i\partial^{\alpha}-\frac{g}{2} a^{\alpha})\psi_+\,+\bar{\psi}_-\bar{\gamma}_{\alpha}(i\partial^{\alpha}+\frac{g}{2} a^{\alpha})\psi_-
-\pi r^2f_{\alpha,\beta}f^{\alpha,\beta}
\end{equation} 
with $f_{\alpha,\beta}=\partial_{\alpha}a_{\beta}-\partial_{\beta}a_{\alpha}$,
where we take only S wave of the gauge fields; $a_0(t,r)\equiv \delta A_0(t,r)$ and $a_1(t,r)\equiv \delta A_r(t,r)$.
The Lagrangian represents a two dimensional Schwinger model with
the boundary condition $\psi_+(r=0)=\bar{\gamma}_0\psi_-(r=0)$. We note that the gauge coupling $g/r$ depends on $r$. 
( The last term in eq(\ref{10}) has the coefficient depending on $r$. It implies the effective gauge coupling being given by $g/r$.  ) 
The model can be exactly solved\cite{rubakov,callan,ezawa}. It has been shown that the charge conservation holds but the chirality conservation is broken.
The restoration of the charge conservation comes from the fact that the deposit of the charge on the monopole after the scattering
is energetically unfavorable; the energy cost for the deposit is proportional to $g^2/r_c$ with small radius e.g. $r_c^{-1}\sim 10^{15}$GeV. 
In QCD, it is inevitable that the production of off diagonal gluons follows the charge non conserved fermion scattering.
Such off diagonal gluons are massive so that the production is energetically unfavorable. Thus, the scattering is not allowed.
Mathematically, the reason why the boundary condition does not work is the presence of 
the strong repulsion when the charged fermion approaches the monopole.
As we note above, the effective gauge coupling grows as $g/r$ when the fermion is close to the monopole. 
Thus, the single fermion can not approach at $r=0$.
A group of fermions with vanishing total charge can approach the monopole.

On the other hand,
the chiral symmetry breaking results from the chiral anomaly. The anomaly arises because 
we take into account the quantum effects of the gauge fields $a_{\alpha}$.
It is effective especially around the monopole; see eq(\ref{3}).
The local presence of the chiral condensate around a monopole has been found \cite{ezawa};
$\langle \bar{\psi}_{\pm}\psi_{\pm}\rangle\sim 1/r$. If we take a chirality non conserved boundary condition in QCD, the chiral
condensate also arises. There are no mechanism for the chirality conservation to be restored.
Therefore, even if we take either the charge non conserved or the chirality non conserved boundary condition in QCD, 
the charge conservation holds but the chirality conservation is broken. It causes the local presence of the chiral condensate around the monopole.
The physical boundary condition imposed on the quarks is such that the quarks conserve their charges but does not conserve their 
chiralities in the scattering with QCD monopoles.

\section{chiral non symmetric pair production}
\label{55}
Up to now, we have shown the local presence of the chiral condensate around a monopole. This is indicative of the fact that the chiral symmetry is broken
by the monopole condensation, although we do not necessarily demonstrate the fact.
Here we show that $\langle dQ_5/dt \rangle\neq 0$ when an external charge is put in the vacuum with the monopole condensation, 
while $\langle dQ_5/dt \rangle=0$ without the monopole condensation.
The non vanishing of $\langle dQ_5/dt \rangle$ implies that the chiral non symmetric spontaneous pair production of the massless fermions arises.

In order to show it we notice the anomaly equation(\ref{3}), 
which describes the chiral non symmetric pair production $dQ_5(\vec{x})/dt\neq 0$ when a charged particle is put at $\vec{x}$
around a single monopole located at $\vec{x}=0$.   
When there are monopoles with their magnetic charges $g_m\eta_i$ at $\vec{x}_i$ ( $ i=1,,,$ ) with $\eta_i=\pm 1$, the anomaly equation is given by

\begin{equation}
\frac{dQ_5(\vec{x})}{dt}=\sum_{i=1,,,}\frac{cg^3g_m\eta_i }{|\vec{x}-\vec{x}_i|}=\int d^3y \frac{cg^3\rho_m(\vec{y})}{|\vec{x}-\vec{y}|} \quad \mbox{with} \quad
\rho_m(\vec{y})\equiv \sum_{i=1,,,}\eta_ig_m\delta(\vec{y}-\vec{x}_i),
\end{equation}  
where $\rho_m$ denotes the magnetic charge density of the monopoles.

In order to discuss the quantum effects of the monopoles, we assume a model of the monopoles, that is, a model
of dual superconductor. The model describes quark confinement by the formation of the color electric flux tube, which
is formed owing to the monopole condensation. The monopole is represented by the complex scalar field $\Phi$ which
minimally couples with dual gauge fields $B_{\mu}$. Namely their coupling is given by the kinetic term of the monopole,
$|D_{\mu}\Phi|^2=|(\partial_{\mu}-ig_mB_{\mu})\Phi|^2$. Then, the magnetic charge density is given by
$\rho_m=g_m\Phi^{\dagger}iD_t\Phi+h.c.$.
We rewrite the anomaly equation in terms of this density $\rho_m$.
Because we wish to obtain the vacuum expectation value $\langle dQ_5/dt\rangle$,
we estimate it using the cluster property such that 

\begin{equation}
\label{12}
\lim_{|\vec{x}|\to \infty}\langle \frac{dQ_5(\vec{x})}{dt}\frac{dQ_5(0)}{dt} \rangle
=\lim_{|\vec{x}|\to \infty}\int d^3yd^3y' \frac{(cg^3)^2\langle\rho_m(\vec{y})\rho_m(\vec{y'})\rangle}{|\vec{x}-\vec{y}||\vec{y}'|}
=\langle dQ_5/dt\rangle^2, 
\end{equation}
with $\rho_m=g_m\Phi^{\dagger}iD_t\Phi+h.c.$,
where the expectation value is taken using the vacuum of the monopoles.
Noting the translational invariance, we rewrite the equation(\ref{12}) with the use of the function 
$f(\frac{\vec{y}-\vec{y}'}{\sqrt{2}})\equiv\langle \rho_m(\vec{y})\rho_m(\vec{y}')\rangle$
and the variables $\vec{y}_{\pm}\equiv(\vec{y}\pm\vec{y}')/\sqrt{2}$,

\begin{eqnarray}
&&\lim_{|\vec{x}|\to \infty}\langle \frac{dQ_5(\vec{x})}{dt}\frac{dQ_5(0)}{dt} \rangle=
\lim_{|\vec{x}|\to \infty}\int d^3y_{+}d^3y_{-} \frac{2(cg^3)^2f(\vec{y}_{-})}{|\vec{y}_{+}+2\vec{y}_{-}-\sqrt{2}\vec{x}||\vec{y}_{+}|} \nonumber \\
&=& \lim_{|\vec{x}|\to \infty}4\pi(cg^3)^2\int_0^{\infty} dy_{+}\int d^3y_{-} 
\frac{\exp(-y_{+}/L)\Big(y_{+}+|2\vec{y}_{-}-\sqrt{2}\vec{x}|-|y_{+}-|2\vec{y}_{-}-\sqrt{2}\vec{x}||\Big)f(\vec{y}_{-})}{|2\vec{y}_{-}-\sqrt{2}\vec{x}|} \nonumber \\
&=& \lim_{|\vec{x}|\to \infty}8\pi L^2(cg^3)^2 \int d^3y_{-} 
\frac{\Big(1-\exp(-|2\vec{y}_{-}-\sqrt{2}\vec{x}|/L)\Big)f(\vec{y}_{-})}{|2\vec{y}_{-}-\sqrt{2}\vec{x}|} \nonumber \\
&\simeq& 8\pi L(cg^3)^2\int d^3y_{-} f(\vec{y}_{-})=8\pi L(cg^3)^2\int d^3y_{-}\langle \rho_m(\sqrt{2}\vec{y}_{-})\rho_m(0)\rangle 
=\frac{4\pi L(cg^3)^2}{\sqrt{2}} \langle Q_m\,\rho_m(0)\rangle ,
\end{eqnarray}
with the magnetic charge $Q_m\equiv \int d^3x \rho(\vec{x})$,
where we have introduced a cut off $L$ in the integration of $y_{+}$ and have taken a limit $L\to \infty$ before taking the limit $|\vec{x}|\to \infty$.
Therefore, the vacuum expectation value $\langle dQ_5/dt\rangle$ is given by

\begin{equation}
\label{14}
 \langle v|\frac{dQ_5}{dt}|v\rangle=\pm \sqrt{\frac{4\pi L(cg^3)^2}{\sqrt{2}} \langle v|Q_m\,\rho_m(0)|v \rangle},
\end{equation}
where we denote the vacuum as $|v\rangle$.

The equation (\ref{14}) implies that when the monopole does not condense $\langle v=0|\Phi|v=0\rangle=0$, the value $\langle v=0|dQ_5/dt|v=0\rangle$ vanishes,
while it does not vanish when the monopole condense $\langle v|\Phi|v\rangle=v \neq 0$.
Actually, when the monopole does not condense, the vacuum is an eigenstate of the magnetic charge $Q_m$. 
We may take the eigenvalue to be zero. Thus $Q_m|v\rangle=0$. It means that $\langle v=0|Q_m\,\rho(0)|v=0 \rangle=0$.
On the other hand, when the monopole condenses, 
the vacuum is not eigenstate of $Q_m$. Thus, $Q_m|v\rangle\neq 0$, although there are no magnetic charge $\langle v| Q_m|v \rangle=0$. 
Generally it means that $\langle v|Q_m\,\rho(0)|v \rangle \neq 0$.
We can explicitly check the result by calculating the expectation $\langle v|\rho(\vec{x})\,\rho(0)|v \rangle$ using the formula

\begin{eqnarray}
0&\neq&\int d^3x\langle v|\rho_m(\vec{x})\,\rho_m(0)|v \rangle \simeq \int d^3x\Big(4v^4g_m^2\langle B_t(\vec{x})B_t(0)\rangle
+16v^2g_m^2\langle\delta\phi(\vec{x})\delta\phi(0)\rangle\langle B_t(\vec{x})B_t(0)\rangle\Big) \quad \mbox{for} \quad \langle\Phi\rangle=v \nonumber \\
0&=&\int d^3x\langle v=0|\rho_m(\vec{x})\,\rho_m(0)|v=0 \rangle\simeq \int d^3x\langle(\Phi^{\dagger}(\vec{x})i\partial_t\Phi(\vec{x})+h.c.)
(\Phi^{\dagger}(0)i\partial_t\Phi(0)+h.c.)\rangle
\quad \mbox{for} \quad \langle\Phi\rangle=0
\label{15}
\end{eqnarray}
with $\Phi=(v+\delta\phi)\exp(i\theta)$ when $\langle\Phi\rangle=v$,
where the angle $\theta$ has been absorbed in $B_t$. We have taken the normal ordering of $\rho_m$ in the evaluation(\ref{15}) in order to have $\langle\rho_m\rangle=0$. 

Therefore, we find that the chiral non symmetric pair production $\langle v|dQ_5/dt|v\rangle\neq 0$
of the fermions arises when a charged particle is put in the vacuum
with the monopole condensation, while the pair production is chiral symmetric $\langle v=0|dQ_5/dt|v=0\rangle=0$ 
when the charged particle is put in the vacuum
with no monopole condensation.
The sign of $\langle v|dQ_5/dt|v\rangle$ is spontaneously determined.

\vspace{0.1cm}
Although we do not still show the standard criterion for the chiral symmetry breaking, i.e. $\langle \bar{q}q \rangle\neq 0$,
our result shows the presence of the phenomena in which the chiral symmetry is spontaneously broken.
Namely, the pair production of the fermions is not chiral symmetric.
It is obvious that
the chiral anomaly plays an essential role for the chiral symmetry breaking in our argument. 

\section{summary}
\label{66}
We have discussed under the assumption of the Abelian dominance in QCD that there exists 
the local chiral condensate around each magnetic monopole. 
Furthermore, we have shown that the chiral non symmetric pair production arises when a charged particle is put in a vacuum
with the monopole condensation. These results indicate
that the chiral symmetry is broken by the monopole condensation. 
Our arguments are based on the Abelian dominance. Recently,
the intimate relation between the QCD monopoles and the chiral symmetry breaking
has been pointed out\cite{hasegawa} beyond the assumption.

In the recent paper\cite{iwazaki}, we have proposed a chiral model of hadrons coupled with the monopoles, in which
the chiral symmetry is spontaneously broken by the monopole condensation.
The one of the monopoles in the model is a color singlet so that it 
is observable. Such a monopole may be observed because we know how the monopole couples with hadrons in the model.
Additionally, we notice the recent papers\cite{heavy} which show that the monopoles
play important roles in producing quark gluon plasma in high energy heavy ion collisions.
In this way, the QCD monopoles have been phenomenologically paid much attention to. 
The present paper would inspire these researches on QCD monopoles.


 \vspace{0.2cm}
The author
expresses thanks to Prof. Kondo, Chiba University and members of KEK for useful comments
and discussions.



\end{document}